\documentclass[12pt]{iopart}
\usepackage{latexsym,amssymb,amsfonts,amsthm,amscd,bbm}
\usepackage[dvipdfm]{graphicx}

\usepackage{cite}
\newcommand{\muB}{\mu_{\rm B}}
\newcommand{\VD}{V_{\rm I}}
\newcommand{\VI}{V_{\rm C}}

\renewcommand{\i}{\ensuremath{{i}}}
\renewcommand{\r}{\ensuremath{\vec{r}}}

\newcommand{\q}{\ensuremath{\vec{q}}}
\renewcommand{\e}[1]{\ensuremath{e^{#1}}}
\newcommand{\dr}{\ensuremath{d^2\vec{r}}}

\begin{document}
%$Revision: 1.93 $

\title[Compressibility stripes for mesoscopic QH
samples]{Compressibility stripes for mesoscopic quantum
Hall samples}

\author{C.\ Sohrmann and R.A.\ R\"{o}mer}

\address{Department of Physics and Centre for Scientific Computing,
University of Warwick, Gibbet Hill Road, Coventry CV4 7AL, UK}
\ead{c.sohrmann@warwick.ac.uk, r.roemer@warwick.ac.uk}

\date{$Revision: 1.93 $, compiled \today}

\begin{abstract}
  We numerically investigate the interplay of disorder and
  electron-electron interactions in the integer quantum Hall effect. In
  particular, we focus on the behaviour of the electronic
  compressibility as a function of magnetic field and electron density.
  We find manifestations of non-linear screening and charging effects
  around integer filling factors, consistent with recent imaging
  experiments.  Our calculations exhibit $g$-factor enhancement as well
  as strong overscreening in the centre of the Landau bands. Even though
  the critical behaviour appears mostly unaffected by interactions,
  important implications for the phase diagram arise.  Our results are
  in very good agreement with the experimental findings and strongly
  support the relevance of electron-electron interactions for
  understanding integer quantum Hall physics.
\end{abstract}

%Uncomment for PACS numbers title message
% Keywords required only for MST, PB, PMB, PM, JOA, JOB?
%\vspace{2pc}
%\noindent{\it Keywords}: Article preparation, IOP journals
% Uncomment for Submitted to journal title message
\submitto{\NJP}
% Comment out if separate title page not required
\maketitle

%\tableofcontents

%%%%%%%%%%%%%%%%%%%%%%%%%%%%%%%%%%%%%%%%%%%%%%%%%%%%%%%%%%%%%%%%%%%%%%%%%%%
\section{Introduction}
\label{sec-introduction}
%%%%%%%%%%%%%%%%%%%%%%%%%%%%%%%%%%%%%%%%%%%%%%%%%%%%%%%%%%%%%%%%%%%%%%%%%%%

The {\em integer}\ quantum Hall effect (IQHE) --- observed in
two-dimensional electron systems (2DES) subject to a strong
perpendicular magnetic field \cite{KliDP80} --- has been well explained
based on single-particle arguments
%\cite{AokA81,Yos02}
\cite{Pra81,ChaP95,JanVFH94,Lau81,Pru84,ThoKNN82,Pru87,ChaC88,CaiR04,KraOK05}.
Crossing the centre of each disorder-broadened Landau level, the Hall
conductivity $\sigma_{xy}$ jumps by $e^2/h$ and the longitudinal
conductivity $\sigma_{xx}$ is finite. The accompanying
localization-delocalization transition is governed by a critical
exponent $\tilde{\nu}=2.34 \pm 0.04$ \cite{HucK90,KocHKP91b}, a
universal quantity independent of microscopic details of electron
motion, the disorder realization, and thus the chosen material. In fact,
it is precisely this astonishing resilience of the quantum Hall (QH)
effect that makes it an ideal metrological standard \cite{KliE85}.

However, recent experiments on mesoscopic MOSFET devices have questioned
the validity of such a simple single-particle picture. Measurements of
the Hall conductance as a function of magnetic field $B$ and gate
voltage \cite{CobBF99} exhibited regular patterns along integer filling
factors. It was argued that these patterns should be attributed to
Coulomb blockade effects.
Similar patterns have been found recently also in measurements of the
electronic compressibility $\kappa$ as a function of $B$ and electron
density $n_{\rm e}$ in the IQHE \cite{IlaMTS04} as well as the fractional
quantum Hall effect (FQHE) \cite{MarIVS04}. From these measurements it
turns out that deep in the localized regime between two Landau levels,
stripes of constant width with particularly small $\kappa$ can be
identified. These stripes consist of a collection of small-$\kappa$
lines, identifiable with localized states, and their number is
independent of $B$. This is inconsistent with a single-particle picture
where one expects a fan-diagram of lines emanating from $(0,0)$ in the
$(B,n_{\rm e})$-plane. The authors argue that their results may be
explained qualitatively by non-linear screening of the impurity charge
density at the Landau level band edges. Clearly such screening effects
--- explained within a Thomas-Fermi approach \cite{IlaMTS04} --- are
beyond a simple non-interacting theory. This immediately raises a
question on the status of the aforementioned universality of the QH
transition \cite{CobBF99,ShaHLT98,BalMB98}, which was obtained largely
within a single-particle approach
%\cite{Pru87,JanVFH94,ChaP95}.
\cite{AokA85,ChaC88,ChaD88,HucK90,LiuS94,Huc95,SinMG00,CaiR05}. Further
qualitative support for the screening arguments where recently given in
\cite{SohR05,StrK06}, where a Hartree-Fock (HF) approach lead to a non-fan
structure for $\kappa$ in the $(B,n_{\rm e})$ diagram; a further
Thomas-Fermi-type investigation showed that charging lines with the
desired slopes can indeed be found within a simple model for the
impurity potentials \cite{PerC05}.

In the present paper, we quantitatively investigate the effects of
Coulomb interactions on the compressibility in the $(B,n_{\rm e})$-plane
within a HF approach. HF accounts for Thomas-Fermi screening effects
while at the same time leading to a critical exponent $\nu$ whose value
is consistent with the results of the non-interacting approaches
\cite{LeeW96,YanMH95}. We find that the observed charging lines in the
compressibility can be well reproduced and that the width of each group of
lines is well estimated by a force balance argument.
Thus we can quantitatively explain the lines and stripes in
the compressibility as a function of $(B,n_{\rm e})$. We note that these
results fully support the qualitative picture proposed in
\cite{IlaMTS04,PerC05,SohR05,StrK06}.

%%%%%%%%%%%%%%%%%%%%%%%%%%%%%%%%%%%%%%%%%%%%%%%%%%%%%%%%%%%%%%%%%%%%%%%%%%%
\section{Formulation of the QH model in Landau basis}
\label{sec-landau}
%%%%%%%%%%%%%%%%%%%%%%%%%%%%%%%%%%%%%%%%%%%%%%%%%%%%%%%%%%%%%%%%%%%%%%%%%%%
In order to model a high-mobility heterostructure in the QH regime, we
consider a 2DES in the $(x,y)$-plane subject to a perpendicular magnetic
field $\vec{B} = B\vec{e}_z$. The system can be described by a
Hamiltonian of the form
\begin{equation}
H^\sigma_{\rm 2DES} =
           h^\sigma + \VI =
           \frac{(\vec{p}-e\vec{A})^2}{2m^*} +
           \frac{\sigma g^* \muB B}{2} +
           \VD(\vec{r}) +
           \VI(\vec{r},\vec{r}'),
\label{eq-hamiltonian}
\end{equation}
where $\sigma = \pm 1$ is a spin degree of freedom, $\VD$ is a smooth
random potential modeling the effect of the electron-impurity
interaction, $\VI$ represents the electron-electron interaction term and
$m^*$, $g^*$, and $\mu_{\rm B}$ are the effective electron mass,
$g$-factor, and Bohr magneton, respectively.
In order to avoid edge effects we impose a torus geometry of size
$L\times L$ onto the system \cite{YosHL83}. The electron-impurity
interaction is modeled by an electrostatic potential due to a remote
impurity density separated from the plane of the 2DES by a spacer-layer
of thickness $d$, as found for instance in modulation-doped GaAs-GaAlAs
heterojunctions. Within the plane of the 2DES, this creates a random,
spatially correlated potential with a typical length scale $d$. We use
$N_{\rm I}$ Gaussian-type "impurities", randomly distributed at
$\vec{r}_s$, with random strengths $w_s \in [-W,W]$, and a fixed width
$d$, such that $ \VD(\vec{r}) = \sum_{s=1}^{N_{\rm I}} \left(w_s/\pi
  d^2\right) \exp[-(\vec{r}-\vec{r}_s)^2/d^2] = \sum_{\vec{q}}
\VD(\vec{q}) \exp(\i\vec{q}\cdot\vec{r}) $ with
\begin{equation}
\VD(\vec{q}) = \sum_{s=1}^{N_{\rm I}} \frac{w_s}{L^2} \exp
\left(-\frac{d^2|\vec{q}|^2}{4}-\i\vec{q}\cdot\vec{r}_s\right),
\end{equation}
where $q_{x,y} = {2\pi j}/{L}$ and $j= -N_\phi, -N_{\phi}-1, \ldots,
N_\phi$. The areal density of impurities therefore is given by $n_{\rm
  I} = N_{\rm I}/L^2$.  The limit $d \rightarrow 0$ yields a potential
of $\delta$-type that would be more adequate for modeling low-mobility
structures.  We will highlight a few differences between the two cases
in the following sections.
%\begin{equation}
%\VD(\vec{r}) = \sum_{s=1}^{S} \sum_{i,j=-\infty}^\infty \frac{V_s}{\pi d^2} \exp \left(-\frac{(\vec{r}-\vec{r}_s+i\vec{e}_xL+j\vec{e}_yL)^2}{d^2}\right).
%\end{equation}
The electron-electron interaction potential
%will be evaluated in Hartree-Fock approximation
%\cite{Aok79,MacG88,MacA86}, resulting in a local Coulomb term and a
%non-local exchange term, but in general
has the form
$
\VI(\vec{r},\vec{r}') =
  \gamma e^2/4\pi\epsilon\epsilon_0|\vec{r}-\vec{r}'| =
  \sum_{\vec{q}} \VI(\vec{q})
  \exp\left[\i\vec{q}\cdot(\vec{r}-\vec{r}')\right],
\label{eq-coulomb-interaction-term}
$
with
\begin{equation}
\VI(\vec{q}) =
\frac{\gamma e^2}{4\pi\epsilon\epsilon_0 l_{\rm c}}\frac{1}{N_\phi |\vec{q}|l_{\rm c}}.
\end{equation}
%\begin{equation}
%\VI(\vec{r},\vec{r}') = \sum_{i,j=-\infty}^\infty \frac{e^2}{4\pi\epsilon\epsilon_0|\vec{r}-\vec{r}'+i\vec{e}_xL+j\vec{e}_yL|},
%\end{equation}
The parameter $\gamma$ will allow us to continually adjust the
interaction strength; $\gamma=1$ corresponds to the bare Coulomb
interaction.  Choosing the vector potential in Landau gauge, $\vec{A} =
Bx\vec{e}_y$, the kinetic part of the Hamiltonian is diagonal in the
Landau functions \cite{LanL81}
\begin{equation}
\varphi_{n,k}(\vec{r}) =
\frac{1}{\sqrt{2^n n! \sqrt{\pi} l_{\rm c} L}}
\exp \left[\i k y-\frac{(x-kl_{\rm c}^2)^2}{2l_{\rm c}^2}\right] H_n\left(\frac{x-kl_{\rm c}^2}{l_{\rm c}}\right),
\end{equation}
where $n$ labels the Landau level index, $k={2\pi j}/{L}$ with $j= 0,
\ldots, N_{\phi}-1$ labels the momentum, $H_{n}(x)$ is the $n$th Hermite
polynomial, and $l_{\rm c} = \sqrt{\hbar/eB}$ the magnetic length.
These functions are extended and $L$-periodic in $y$-direction and
localized in $x$ direction.\\
For the system's many-body state, $|\Phi\rangle$, we use the usual
ansatz \cite{Aok79,MacA86} of an anti-symmetrized product of single
particle wave-functions $\psi_\alpha^\sigma(\vec{r})$ (Slater
determinant), which we choose as a linear combination of Landau states
\begin{equation}
  \psi_\alpha^\sigma(\vec{r}) = \sum_{n=0}^{N_{\rm LL}-1}
  \sum_{k=0}^{N_\phi-1} \vec{C}^{\alpha,\sigma}_{n,k} \chi_{n,k}(\vec{r}),
\label{eq-ansatzwf}
\end{equation}
with $N_{\rm LL}$ being the number of Landau levels and the periodic
Landau functions
\begin{equation}
\chi_{n,k}(\vec{r}) = \langle \vec{r}|nk\rangle =
\sum_{j=-\infty}^\infty \varphi_{n,k+j L/l_{\rm c}^2}(\vec{r}),
\end{equation}
in order to meet the boundary conditions. The number of flux quanta
piercing the 2DES of size $L\times L$ is given by $N_\phi=L^2/2\pi
l_{\rm c}^2$, yielding a total number of $M = N_{\rm LL} N_\phi$ states
per spin direction. The filling of the system is characterized by the
filling factor $\nu = N_{\rm e}/N_\phi$, with $N_{\rm e}$ being the
number of electrons in the system and areal density $n_{\rm e} = N_{\rm
  e}/L^2$. The total Landau level density is given by $n_0 = eB/h$.

%%%%%%%%%%%%%%%%%%%%%%%%%%%%%%%%%%%%%%%%%%%%%%%%%%%%%%%%%%%%%%%%%%%%%%%%%%%
\section{Hartree-Fock equation in the Landau basis and its numerical solution}
\label{sec-hartree-fock}
%%%%%%%%%%%%%%%%%%%%%%%%%%%%%%%%%%%%%%%%%%%%%%%%%%%%%%%%%%%%%%%%%%%%%%%%%%%

We are left with finding the correct expansion coefficients
$\vec{C}^{\alpha,\sigma}_{n,k}$ \cite{Aok79,YosF79,MacA86,MacG88}. The
Hamiltonian is represented in matrix form using the periodic Landau states
$|nk\rangle$ and we have
\begin{eqnarray}
\mathbf{H}^\sigma_{n,k;n',k'}
&= &\langle nk |H^\sigma_{\rm 2DES}|n'k'\rangle \nonumber \\
&= &\left(n+\frac{1}{2}+\frac{\sigma g^*}{4}\frac{m^*}{m_{\rm e}}\right)
     \hbar\omega_{\rm c}\delta_{n,n'}\delta_{k,k'} +
 \mathbf{V}_{n,k;n',k'} + \mathbf{F}^\sigma_{n,k;n',k'}\quad,
\end{eqnarray}
with the cyclotron energy $\hbar\omega_{\rm c} = \hbar eB/m^*$.  The
disorder matrix elements are given by $\mathbf{V}_{n,k;n',k'} =
\sum_{\vec{q}} \VD(\vec{q}) S_{n,k;n',k'}(\vec{q})$, where mixing of
Landau levels is included. The explicit form of the plane wave matrix
elements $S_{n,k;n',k'}(\vec{q}) = \langle nk
|\exp(\i\vec{q}\cdot{\vec{r}})|n'k'\rangle$ is computed in the Appendix.
The elements of the Fock matrix $\mathbf{F}$ are
\begin{equation}
\mathbf{F}^\sigma_{n,k;n',k'} =
 \sum_{\sigma'}\sum_{l,m,l',m'}
 \left( G_{n,k;n',k'}^{m,l;m',l'}-\delta_{\sigma,\sigma'}G_{n,k;m',l'}^{m,l;n',k'}
 \right)  \mathbf{D}^{\sigma'}_{m,l;m',l'}\quad,
\label{eq-fockmatrix}
\end{equation}
where the first term is the Hartree and the second the Fock contribution.
The bielectronic integrals
$
G_{n,k;n',k'}^{m,l;m',l'} = \sum_{\vec{q} \neq 0} \VI(\vec{q})
  S_{n,k;n',k'}(\vec{q}) %\langle nk|\e{\i\vec{r}\cdot\vec{q}}|n'k'\rangle
  S_{m,l;m',l'}(-\vec{q}) %\langle ml|\e{-\i\vec{r}\cdot\vec{q}}|m'l'\rangle
%\label{eq-bielectronic}
  $ can be further simplified as given in the Appendix. A homogeneous,
  positive background is assumed that neutralizes the charge of the
  electrons and thereby prevents the Coulomb term from diverging as
  $|\vec{q}| \rightarrow 0$.  In fact, this interaction with the
  background can be shown to cancel with the term $|\vec{q}|=0$ in
  $\mathbf{F}$ up to a contribution of the order of $L^{-1}$ due to the
  finite system size \cite{BacH01}.  The density matrix is given by
\begin{equation}
  \mathbf{D}^\sigma_{m,l;m',l'} =
  \sum_{\alpha=1}^{M}
  f(\epsilon^\sigma_\alpha) \left(\mathbf{C}^{\alpha\sigma}_{m,l}\right)^*
  \mathbf{C}^{\alpha,\sigma}_{m',l'}\quad,
\label{eq-densitymatrix}
\end{equation}
with
%\begin{equation}
$\Tr(\mathbf{D}) = N_{\rm e}$
%\end{equation}
and
%\begin{equation}
$\mathbf{D}^\sigma \mathbf{D}^\sigma = \mathbf{D}^\sigma$.
%\end{equation}
Here $f(\epsilon)$ denotes the Fermi function.  The total energy $E_{\rm
  tot}$ in terms of the above matrices is given as
\begin{equation}
E_{\rm tot}  %= \Tr(\mathbf{h}\mathbf{D}) + \frac{1}{2}\Tr(\mathbf{F}\mathbf{D})
             =\Tr(\mathbf{h}\mathbf{D}+\frac{1}{2}\mathbf{F}\mathbf{D})
             = \frac{1}{2}\sum_\sigma \sum_{n,k;n',k'}
             \left(2 \mathbf{h}^\sigma_{n,k;n',k'}+\mathbf{F}^\sigma_{n,k;n',k'}\right)
             \mathbf{D}^\sigma_{n,k;n',k'}.
\label{eqn-diffEtot}
\end{equation}
A variational minimization of $\langle\Psi|H_{\rm 2DES}|\Psi\rangle$
with respect to the coefficients yields the Hartree-Fock-Roothaan
equation \cite{Roo51}, a self-consistent eigenvalue problem which in
compact form can be written as
\begin{equation}
\mathbf{H}^\sigma\mathbf{C}^\sigma = \mathbf{C}^\sigma\mathbf{E}^\sigma,
\label{eq-hfe}
\end{equation}
with
%\begin{equation}
$\mathbf{C}^\sigma = (\vec{C}^{\sigma}_1,\dots,\vec{C}^{\sigma}_M)$
%\end{equation}
being the matrix of eigenvectors and
%\begin{equation}
$\mathbf{E}^\sigma =
\mbox{diag}(\epsilon^\sigma_1,...,\epsilon^\sigma_M)$
%\label{eq-eigenvalues}
%\end{equation}
the diagonal matrix of the eigenvalues $\epsilon^\sigma_1 \le
\epsilon^\sigma_2 \le \dots \le \epsilon^\sigma_M$ \cite{overlapmatrix}.\\
Following the aufbau principle \cite{CanB00}, the density matrix is
constructed starting from the energetically lowest lying state up to the
Fermi level $\epsilon_{\rm F}$. In our calculations, we keep $N_{\rm e}$
fixed and compute $\epsilon_{\rm F}$ as the energy of the highest
occupied state afterwards.
Since the Fock matrix depends on the density matrix, which in turn
depends on the full solution of problem, (\ref{eq-hfe}) has to be
calculated self-consistently which is numerically quite challenging.  In
the first step we use the solution of the non-interacting Hamiltonian
$\mathbf{h}^\sigma=\langle nk |h^\sigma|n'k'\rangle$ as a starting guess
for the coefficients $\mathbf{C}^{\sigma}$. From this solution,
$\mathbf{C}^{(0)}$, we construct the density and Fock matrices and
finally the full Hamiltonian. Diagonalization yields an improved
solution, $\mathbf{C}^{(1)}$. The process continues until convergence of
the density matrix has been achieved. In practice, we compute the norm
of the difference between successive density matrices $
||\mathbf{D}^{(n+1)}-\mathbf{D}^{(n)}|| < \varepsilon.  $ Here
$||\cdot||$ denotes the Hilbert-Schmidt norm defined as $||\mathbf{A}||
= \Tr(\mathbf{A}\mathbf{A}^*)^{1/2}$.

However, convergence of this Roothaan algorithm \cite{Roo51} is not
always assured. Especially for fillings close to zero or an integer
value it may run into an oscillating limit cycle. Subtracting a small
multiple of the density matrix, $b\mathbf{D}^\sigma$, from the
Hamiltonian $\mathbf{H}^\sigma = \mathbf{h}^\sigma + \mathbf{F}^\sigma$
favours already occupied states, thereby suppressing such oscillations.
The eigenvalues shift by $\epsilon^\sigma_i \rightarrow
\epsilon^\sigma_i - b \quad \forall \quad \epsilon^\sigma_i \le
\epsilon_{\rm F}$, but eigenstates are not affected. This algorithm is
known as the level shifting algorithm \cite{SauH73} and $b$ is the
level-shift parameter.  We choose $b \approx 1.76 e^2/L$ which is
exactly the contribution of the neglected $V_{\rm C}(\vec{q}=0)$ term
that does not cancel with the background charge due to the finite system
size. This prevents the occupied and unoccupied states from mixing and
stabilizes the convergence of the algorithm, although at the cost of
slower convergence.
A further improvement of convergence is achieved by constructing the new
density matrix ${\mathbf{D}}^{(n+1)}$ as a mixture of the updated and
the previous one, i.e.\ $\tilde{\mathbf{D}}^{(n+1)} = \lambda
\mathbf{D}^{(n+1)} + (1-\lambda) \mathbf{D}^{(n)}$, with $\lambda
\in{[0,1]}$. We adjust $\lambda$ in each step automatically such that
the fastest global convergence is guaranteed \cite{CanB00}.

In each HF step, assembling the dense Fock matrix $\mathbf{F}^\sigma$
scales as $\mathcal{O}(N_{\rm LL}^4 N_\phi^3)$ and is clearly very
time-consuming. An improved scheme, even though generally possible, is
of little advantage since the diagonalization is of similar complexity.
For the calculation of a particular disorder configuration and magnetic
field, a self-consistent run has to be made for each of the $M$ possible
filling factors. Hence, the complexity of a complete HF calculation is
of the order $\mathcal{O}(2 K N_{\rm LL}^5 N_\phi^4)$ with $K$ the
number of iterations until convergence. The dependence on the system
size is therefore $\mathcal{O}(L^8)$. For system sizes of $L \sim
300$nm, we find $K \sim 100-1000$. In all results present here,
convergence of the HF scheme is assumed for $\varepsilon \le 10^{-6}$.

%%%%%%%%%%%%%%%%%%%%%%%%%%%%%%%%%%%%%%%%%%%%%%%%%%%%%%%%%%%%%%%%%%%%%%%%%%%
\section{Chemical potential and electronic compressibility}
\label{sec-chem-pot-compress}
%%%%%%%%%%%%%%%%%%%%%%%%%%%%%%%%%%%%%%%%%%%%%%%%%%%%%%%%%%%%%%%%%%%%%%%%%%%
The electronic compressibility $\kappa=({\partial n_{\rm e}}/{\partial
  \mu})/{n_{\rm e}^2}$ reflects the ability of the 2DES to absorb
electrons when changing the chemical potential. With $\mu = {\partial
  E_{\rm tot}}/{\partial N_{\rm e}}$, we find ${\partial \mu}/{\partial
  n_{\rm e}} = L^2 ({\partial^2 E_{\rm tot}}/{\partial N_{\rm e}^2})$.
Hence, for finite sample calculations, we can obtain $\kappa$ from
$E_{\rm tot}(N_{\rm e})$ using
\begin{equation}
  \frac{\partial \mu}{\partial n_{\rm e}} \approx L^2
  \left[ E_{\rm tot}(N_{\rm e}+1) - 2 E_{\rm tot}(N_{\rm e}) + E_{\rm tot}(N_{\rm e}-1) \right].
\end{equation}
Alternatively, at $T=0$, we can compute the change in the chemical
potential for $N_{\rm e}$ electrons by noting that the Fermi energy
$\epsilon_{\rm F}(N_{\rm {\rm e}})= \mu(N_{\rm {\rm e}})$. Thus we immediately have
\begin{equation}
  \frac{\partial \mu}{\partial n_{\rm {\rm e}}} =
  L^2\left[\epsilon_{\rm F}(N+1)-\epsilon_{\rm F}(N)\right].
\label{eq-diff-fermi}
\end{equation}
This turns out to be numerically more stable than (\ref{eqn-diffEtot})
and shall be used in the following.

When the Fermi energy lies in a region of highly localized states, it
takes a considerable energy to accommodate another electron, and thus
the compressibility is low. On the other hand, in a region of
delocalized states a newly added electron is much more easily absorbed
and $\kappa$ is high. For a non-interacting system, $\kappa$ is
proportional to the tunneling density of states (DOS) \cite{Mah00} and
expected to exhibit a fan-like structure in the $(B,n_{\rm e})$ diagram.
In particular, the resonances in $\kappa$ need not align with slopes
equal to integer filling factors \cite{JaiK88}. In the interacting case,
$\kappa$ is proportional to the thermodynamic density of states (TDOS)
and the inverse screening length.  We note that experimentally the
change of the chemical potential is detected when changing the back gate
voltage and hence the electron density at constant $B$
\cite{IlaMTS04,MarIVS04}.

%%%%%%%%%%%%%%%%%%%%%%%%%%%%%%%%%%%%%%%%%%%%%%%%%%%%%%%%%%%%%%%%%%%%%%%%%%%
\section{Sample mobility and DOS}
\label{sec-mobility-dos}
%%%%%%%%%%%%%%%%%%%%%%%%%%%%%%%%%%%%%%%%%%%%%%%%%%%%%%%%%%%%%%%%%%%%%%%%%%%
In order to tune our parameters for the electron-impurity interaction to
the experiment, we estimate the zero field mobility, defined as
$\mu_{\rm 0} = e \tau/m^*$, with $\tau$ being the transport scattering
time \cite{AshM76,Mah00}. For a short-range $\delta$-impurity potential,
$\tau$ is identical to the single-particle momentum relaxation time,
$\tau_{\rm s}$, which determines the level broadening, $(\Gamma/2)^2 =
\hbar\omega_{\rm c}\hbar/2\pi\tau_{\rm s}$ \cite{AndFS82}. For
long-range potential, however, these two times can be very different
\cite{DasS85} and knowledge about the level broadening does not
necessarily imply knowledge about the mobility and vice versa. In fact,
for a smooth potential with $d \gg l_{\rm c}$ we have $(\Gamma/2)^2 =
\left<\left[\VD(\vec{r})-
    \left<\VD(\vec{r})\right>_{\vec{r}}\right]^2\right>_{\vec{r}}$ ,
which does not depend on $B$. In that case, we can determine the
mobility from the transport cross-section calculated in Born
approximation \cite{LanL81,AndFS82}. Since the transport scattering time
is momentum dependent, we take the low temperature limit, where the
relevant scattering time is the one for electrons having Fermi momentum
$k_{\rm F} = (2\pi n_{\rm e})^{1/2}$. With a radially symmetric
electron-impurity interaction potential (for a single scatterer),
$u(\vec{r})$, we obtain for the transport scattering time
\begin{equation}
  \tau^{-1} =
  \frac{n_{\rm I}m^*}{2\pi\hbar^3}
  \int_0^{2\pi} d\theta \left[1-\cos(\theta)\right]
  \left|\tilde{u}(2k_{\rm F}\sin(\theta/2))\right|^2,
\label{eq-scattering-time}
\end{equation}
with the Fourier transform $u(\vec{k}) = \int d^2\vec{r}
u(\vec{r})\exp(-i\vec{r}\cdot\vec{k})$ and $\tilde{u}(k) =
u(|\vec{k}|)$. In case of $\delta$-interaction and uniformly distributed
strengths, we simply have $\tilde{u}(k) = W/\sqrt{3}$ and the
$\cos(\theta)$ term in (\ref{eq-scattering-time}) vanishes. Without it,
(\ref{eq-scattering-time}) becomes the expression for $\tau_{\rm s}$,
which proves the equivalence of $\tau$ and $\tau_{\rm s}$ for
short-range potentials. For long-range potentials, however, forward
(small $\theta$) scattering receives little weight since it hardly
impairs the electron movement and $\tau/\tau_{\rm s} \gg 1$. In order to
model a situation comparable to the experiments of \cite{IlaMTS04}, we
use material parameters for GaAs ($g^* = -0.44$ for the effective
$g$-factor, $m^* = 0.067m_{\rm e}$ for the effective mass) and impurity
parameters of $W/$nm$^2 \simeq 4$eV with a concentration of $n_{\rm I} =
3.2\cdot10^{11}$cm$^{-2}$ (e.g.~$N_{\rm I}=288$ for $L=300{\rm nm}$ or
$N_{\rm I}=392$ for $L=350{\rm nm}$). Assuming $n_{\rm e} \approx n_{\rm
  I}$, for the $\delta$-potential this yields a mobility of $\mu_0
\simeq 10^3 \mbox{cm}^2/\mbox{Vs}$, whereas for $d = 40$nm ($\approx$
spacer layer thickness) we get $\mu_0 \simeq 10^6 \mbox{cm}^2/\mbox{Vs}$
--- a value which is reasonable for a high mobility sample such as a
GaAs-GaAlAs heterojunction.

\begin{figure}[h]
  \center
  \includegraphics[width=\textwidth]{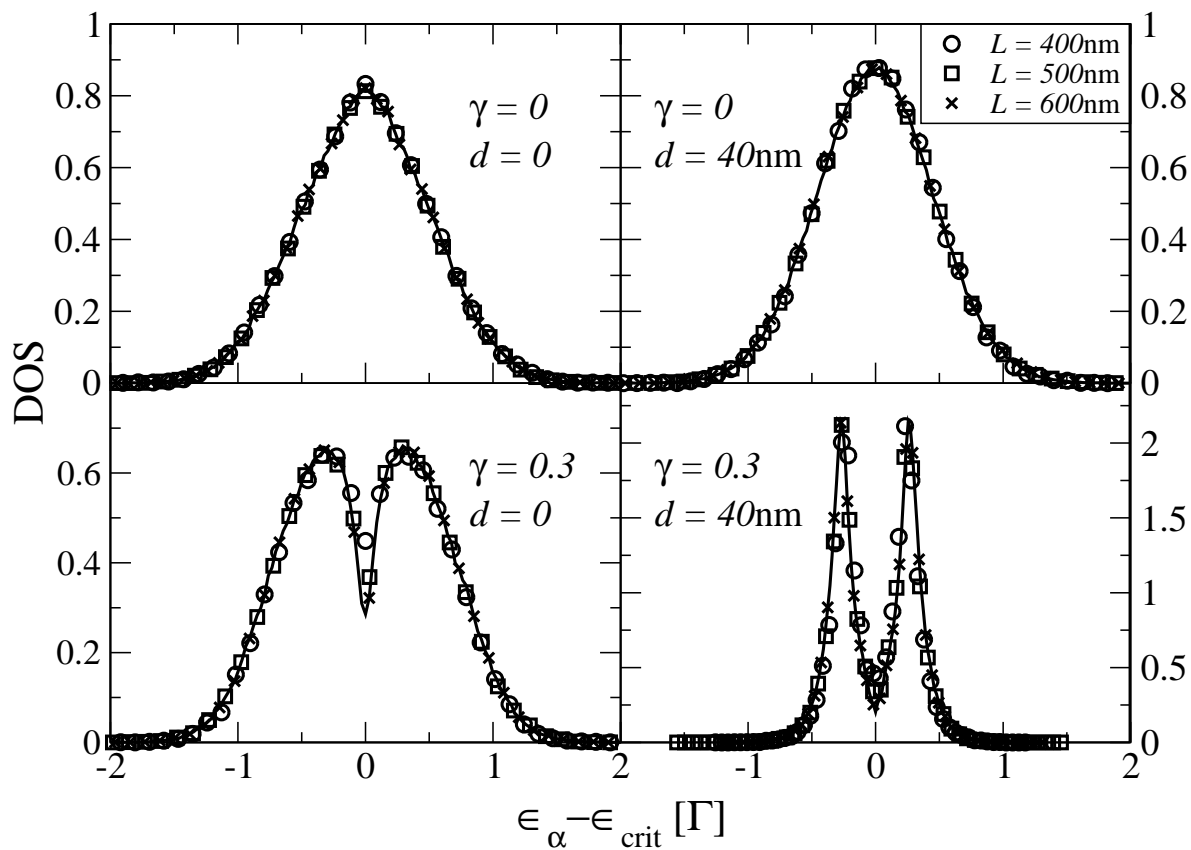}
  \caption{ \label{fig-dos-no-int} DOS at $B=3$T for the lowest Landau
    level in a non-interacting (top row, $\gamma = 0$) and a
    HF-interacting (bottom row, $\gamma = 0.3$ at $\nu=1/2$) QH system
    for $3$ system sizes. The left column shows results for
    $\delta$-type impurities ($d=0$) with $W/\mbox{nm}^2 = 2$eV, the
    right column corresponds to Gaussian-type impurities with $d=40$nm
    and $W/\mbox{nm}^2 = 4$eV ($W/d^2=2.5$meV). The results in all cases
    are averaged over at least $1000$ samples. Error bars are less than
    the symbol sizes. Note the strong Coulomb reduction of the DOS at
    the critical energy ($\epsilon_\alpha = \epsilon_{\rm crit}$) in the
    interacting systems.}
  \label{fig-DOS}
\end{figure}

We have calculated the DOS in the lowest Landau level by averaging over
at least $1000$ samples in the non-interacting and the interacting
system. In Figure \ref{fig-DOS} we show results for Gaussian and
$\delta$-impurities for three system sizes $L = 400, 500$, and $600$nm
for the non-interacting case as well as the interacting case. Within the
accuracy of the calculation, we find the DOS to be independent of the
system size, irrespective of interactions, as of course expected for our
static disorder model \cite{systemsize}.  Furthermore, in the
interacting case, we find a strong suppression of the DOS at the Fermi
level. The formation of this Coulomb gap and its non-criticality has
been studied previously \cite{YanM93,LeeW96,YanMH95,BacH01}. In case of
Gaussian-type impurities we also observe a strong reduction of the band
broadening due to screening of the impurity potential, while the
bandwidth in systems with $\delta$-type impurities is hardly affected by
interactions.
%
%The interacting results are obtained with a level shift-parameter as
%given by (\ref{eq-level-shift-parameter}). The DOS has been calculated
%using $b = 0$ in order to avoid the additional gap at the Fermi level.
%This, however, brings about a finite size effect for smaller systems.
%There is a peak visible exactly at $\nu = 0.5$, with presumably no
%physical origin.

%%%%%%%%%%%%%%%%%%%%%%%%%%%%%%%%%%%%%%%%%%%%%%%%%%%%%%%%%%%%%%%%%%%%%%%%%%%
\section{Scaling of the participation ratio}
\label{sec-part-ratio-scaling}
%%%%%%%%%%%%%%%%%%%%%%%%%%%%%%%%%%%%%%%%%%%%%%%%%%%%%%%%%%%%%%%%%%%%%%%%%%%

\begin{figure}[h]
  \center
  \includegraphics[width=\textwidth]{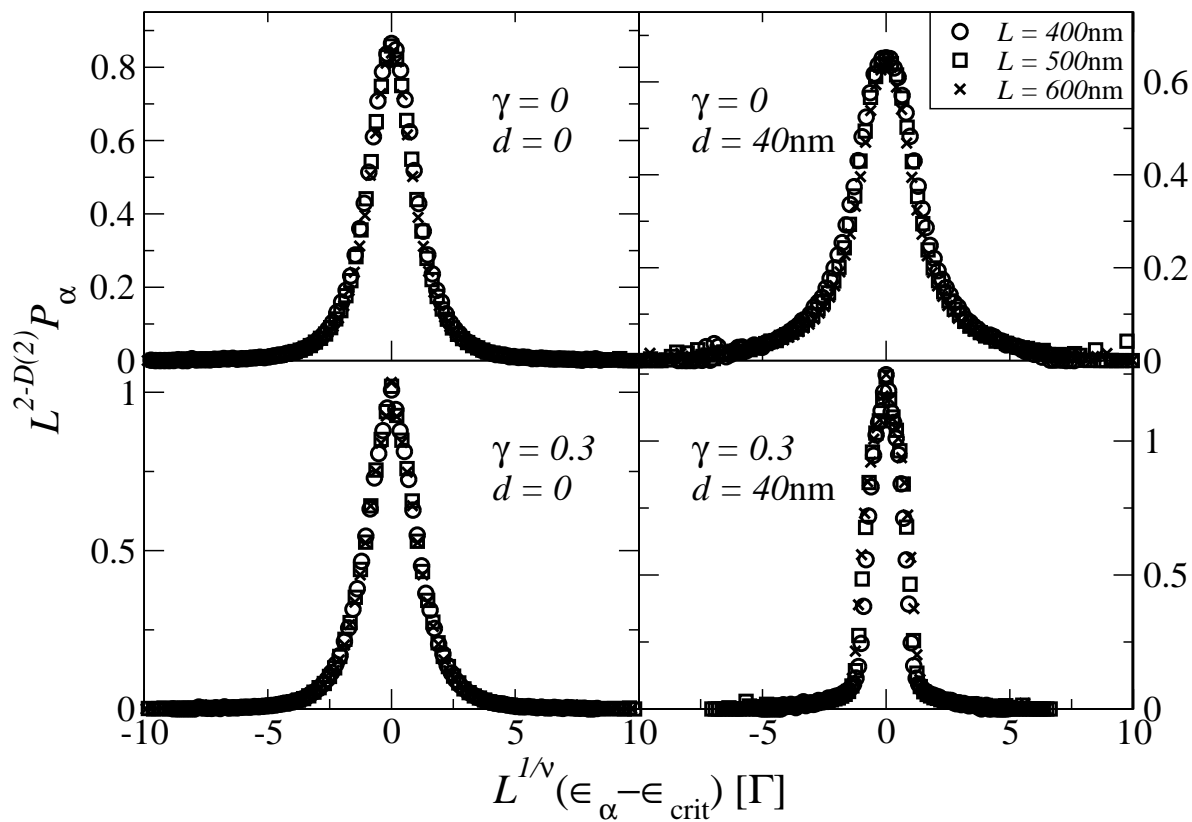}

  \caption{\label{fig-part-ratio} Scaling functions of the participation
    ratio $P_\alpha$ at $B=3$T for the non-interacting (top row, $\gamma
    = 0$) and the HF-interacting (bottom row, $\gamma = 0.3$ at
    $\nu=1/2$) systems averaged over at least $1000$ samples and using
    $D(2)=1.62$, $\tilde{\nu}=2.34$. The left column shows results for
    $\delta$-type impurities ($d=0$), the right column corresponds to
    Gaussian-type impurities with $d=40$nm. Values for $L$ have been
    scaled by the magnetic length. Fluctuations in the tails are due to
    a smaller number of data points.}
\end{figure}

\begin{figure}[h]
  \center
  \includegraphics[width=0.95\textwidth]{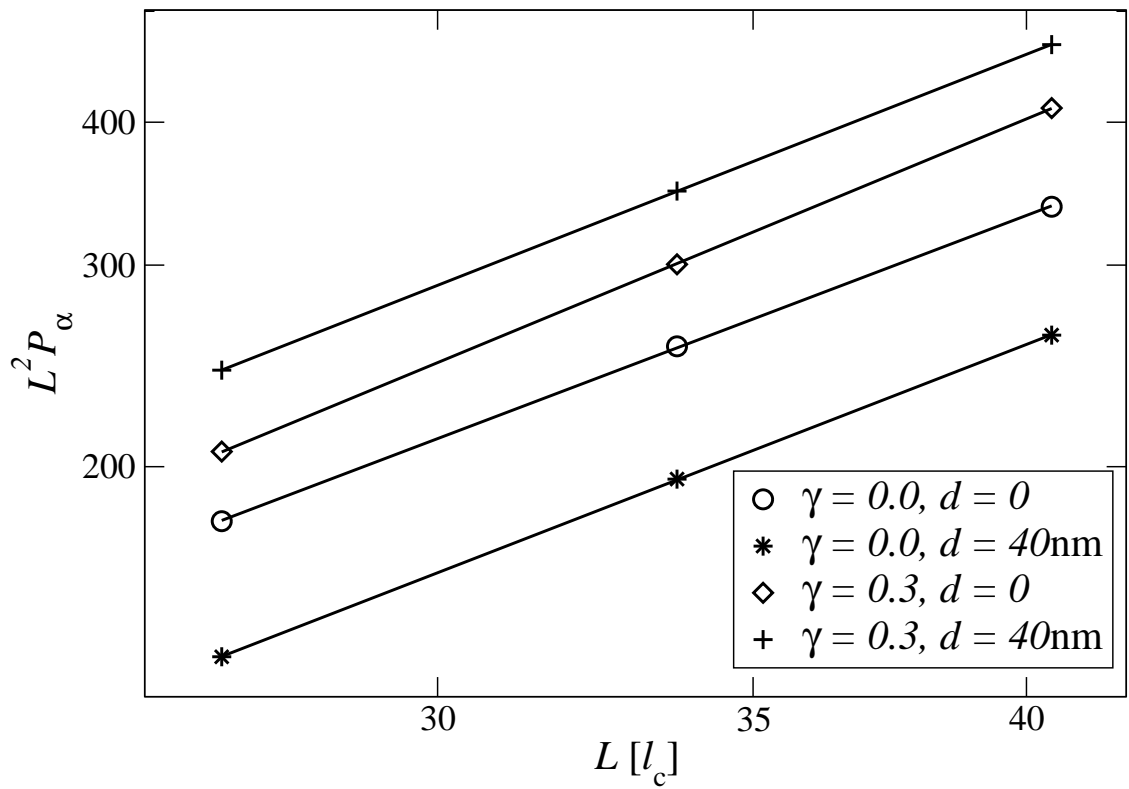}
  \caption{\label{fig-part-ratio-scaling-D2} Power-law fit of the system
    size dependence of $P_{\alpha}$ at $B=3$T according to
    (\ref{eq-part-ratio-scaling}) for non-interacting and HF-interacting
    systems around $\epsilon_\alpha=\epsilon_{\rm crit}$. $\circ$ and
    $\diamond$ denote $\delta$-type impurities ($d=0$) whereas $*$ and
    $+$ show results for Gaussian-type impurities with $d=40$nm. The
    error bars for the data points are smaller than the
    symbol sizes. The $D(2)$ values are $1.57\pm 0.2$ ($\gamma=0$,
    $d=0$), $1.6\pm 0.1$ ($\gamma=0$, $d=40$nm), $1.7\pm 0.1$
    ($\gamma=0.3$, $d=0$), $1.62\pm 0.05$ ($\gamma=0.3$, $d=40$nm).
    Values for $L$ are scaled by the magnetic length.}
\end{figure}

The participation ratio $P_{\alpha}$ is defined as the inverse of the
variance of the charge density in the state $\alpha$,
\begin{equation}
  P_\alpha = \left( L^2\int\dr|\psi_\alpha(\vec{r})|^4 \right)^{-1} \quad .
\end{equation}
Large values of $P_{\alpha}$ correspond to spatially extended states,
while low values indicate a confined state \cite{Aok83,Aok86}. This is
intuitively understood by the fact that the density of an extended state
varies much less over space than a highly localized one. Thus
$P_{\alpha}$ is a measure of the degree of localization
\cite{partratiocare} and may be computed in our model as
\begin{equation}
P_\alpha = \frac{l_{\rm c}^2}{L^2}\sum_{n,n',m,m' \atop k,k',l,l'}\sum_{\vec{q}}
                            (\mathbf{C}^{\alpha}_{n,k})^*
                            \mathbf{C}^{\alpha}_{n',k'}
                            (\mathbf{C}^{\alpha}_{m,l})^*
                            \mathbf{C}^{\alpha}_{m',l'}
                            S_{n,k;n',k'}(\vec{q})
                            S_{m,l;m',l'}(-\vec{q}).
\end{equation}
It has been shown that unscreened HF-interactions do not alter the
critical exponent $\tilde{\nu}$ while renormalizing the dynamical
scaling exponent to $z=1$ \cite{YanMH95,LeeW96,BacH98}. As a check to
our HF results, we calculate $P_{\alpha}$ of spinless electrons in the
lowest Landau level for the same samples as in Section
\ref{sec-mobility-dos}. The participation ratio is expected to obey the
single parameter scaling form \cite{BauCS90}
\begin{equation}
P_\alpha = L^{D(2)-2}\Pi\left(L^{1/\tilde{\nu}}|\epsilon_\alpha-\epsilon_{\rm crit}|\right),
\label{eq-part-ratio-scaling}
\end{equation}
with the anomalous diffusion coefficient $D(2) \approx 1.6$ --- related
to the multifractal character of the critical states
\cite{SouE84,Jan94a,HuckS94} --- and the critical exponent $\tilde{\nu}
\approx 2.3$ \cite{HucK90}. Figure \ref{fig-part-ratio} shows the
scaling function for the non-interacting and an interacting system at
filling factor $\nu = 1/2$.
The scaling function collapses reasonably well onto a single curve for
both, non-interacting and HF-interacting systems. We find $D(2) = 1.62
\pm 0.10$ as typical average over both, HF- and non-interacting systems
as shown in Figure \ref{fig-part-ratio-scaling-D2}.
This demonstrates the irrelevance of interactions and the type of
disorder for the multifractal dimension of the critical state, in very
good agreement with previous results \cite{HuckS94,YanMH95,StrK06}.
%
%We remark that while the scaling in the centre of the band is very good,
%it becomes initially worse for $|E|\approx 5$ and then gets better
%around the band edges.
%
A similar fit in the tails of $P_{\alpha}$ is numerically less accurate
but still yields estimates for $\tilde{\nu}$ between $2$ and $2.4$,
compatible with the expected value $2.34 \pm 0.04$ \cite{HucK90,Huc95}.

%%%%%%%%%%%%%%%%%%%%%%%%%%%%%%%%%%%%%%%%%%%%%%%%%%%%%%%%%%%%%%%%%%%%%%%%%%%
\section{Charging lines and stripes in the electronic compressibility}
\label{sec-compress-stripes}
%%%%%%%%%%%%%%%%%%%%%%%%%%%%%%%%%%%%%%%%%%%%%%%%%%%%%%%%%%%%%%%%%%%%%%%%%%%

Let us first briefly recall the experimental results of
\cite{IlaMTS04,MarIVS04} most relevant to the present investigation. The
compressibilities in the $(B,n_{\rm e})$-plane (i) exhibit only little
variation in regions close to the QH transitions at half-integer filling
factors, but (ii) show a strong variation between Landau bands at
integer fillings which by virtue of the relation $n_{\rm e} = \nu eB/h$
correspond to lines of constant slope.
Furthermore, (iii) these regions of strong variation seem to have a
width which is $B$ and Landau level index independent and (iv) within
these stripes, thin lines of equal slope $j eB/h$, $j=0, 1, \ldots$ can
be identified. In what follows, we have calculated the electronic
compressibility as outlined in Section \ref{sec-chem-pot-compress} in
the lowest two, spin-split Landau levels for a sample of size $L=300$nm
at magnetic fields between $B=0.2$T and $B=6$T.

%%%%%%%%%%%%%%%%%%%%%%%%%%%%%%%%%%%%%%%%%%%%%%%%%%%%%%%%%%%%%%%%%%%%%%%%%%%
\subsection{Compressibilities for the non-interacting system}
\label{sec-compress-non}
%%%%%%%%%%%%%%%%%%%%%%%%%%%%%%%%%%%%%%%%%%%%%%%%%%%%%%%%%%%%%%%%%%%%%%%%%%%

\begin{figure}[h]
  \center
  \includegraphics[width=\textwidth]{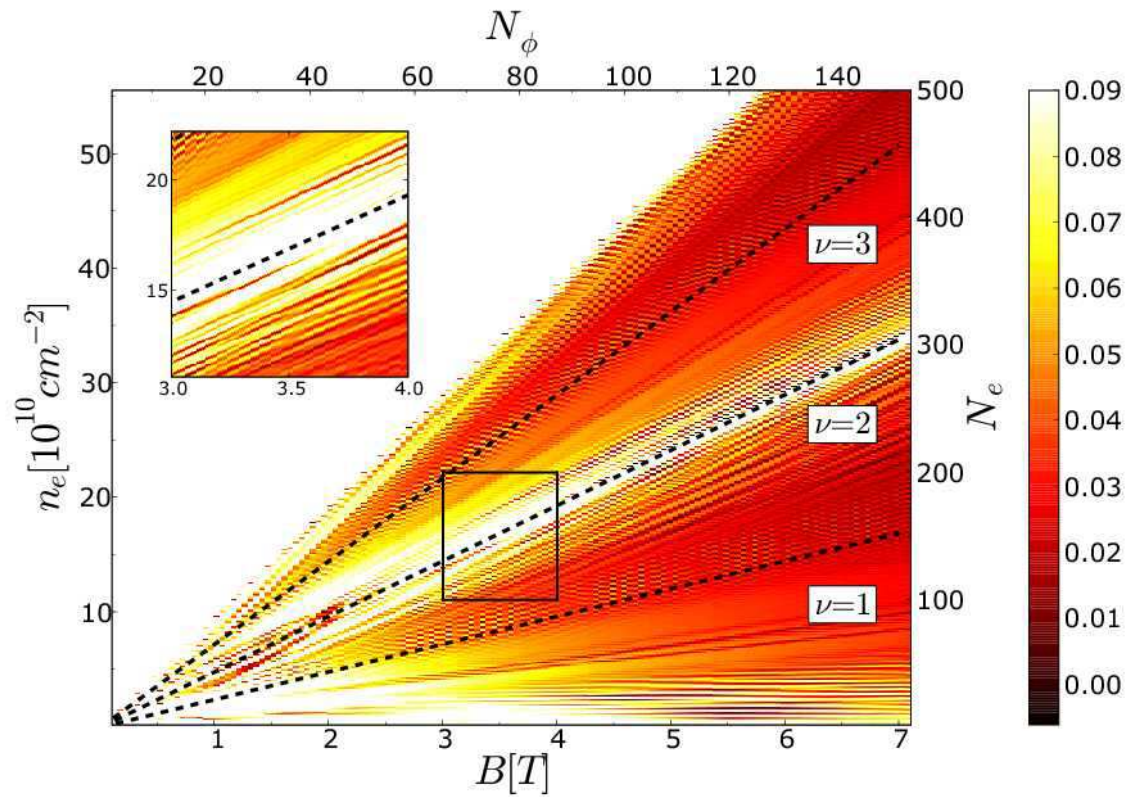}
  \caption{Inverse electronic compressibility $\kappa^{-1}$ for a
    non-interacting system of size $L=300$nm with disorder strength
    $W/d^2 = 2.5$meV in the $(B,n_{\rm e})$-plane. The color scale spans
    two standard deviations around the average of $\kappa^{-1}$. The
    inset shows more detailed results for the region marked by a black
    rectangle.}
  \label{fig-non-compress-1}
\end{figure}

Figure \ref{fig-non-compress-1} shows our results for the {\em inverse}
compressibility $\kappa^{-1}$ of non-interacting ($\gamma=0$) electrons
in the two lowest orbital Landau levels, including spin.
Darker areas in the plot reflect states of higher compressibility, hence
a more delocalized regime. Lighter areas are more strongly localized
states. Due to the weak Zeemann splitting, we do not observe the two
spin bands separately. Rather, both bands remain nearly degenerate and
lie almost on top of each other. Hence, we only find a single, very
strongly incompressible region between the first and the second orbital
Landau level at $\nu = 2$. This broad line is due to the band gap and
the highly localized states at the band edges. Other less pronounced
lines are visible along different filling factors, seemingly mostly
emanating from $(0,0)$. Some lines even appear to have a varying slope
as shown in the inset of Figure \ref{fig-non-compress-1}. We interpret
these features as the aforementioned fingerprints of scattering
resonances in the disorder potential \cite{JaiK88} which do not
necessarily need to align with constant filling factors. Moreover, we
clearly observe an increasing number of those lines with increasing
magnetic field. At $\nu=0$ and $4$, the compressibility is again low.

%%%%%%%%%%%%%%%%%%%%%%%%%%%%%%%%%%%%%%%%%%%%%%%%%%%%%%%%%%%%%%%%%%%%%%%%%%%
\subsection{Effects of the Hartree-Fock interaction}
\label{sec-compress-HF}
%%%%%%%%%%%%%%%%%%%%%%%%%%%%%%%%%%%%%%%%%%%%%%%%%%%%%%%%%%%%%%%%%%%%%%%%%%%

We next include interaction with $\gamma=0.3$. This is not yet the
full $\gamma=1$ Coulomb term, but the results are numerically more
stable while at the same time not being dramatically different from
$\gamma=1$. Furthermore, $\gamma < 1$ is essentially equivalent
to increased disorder with the full Coulomb interaction present.

\begin{figure}[h]
  \center
  \includegraphics[width=\textwidth]{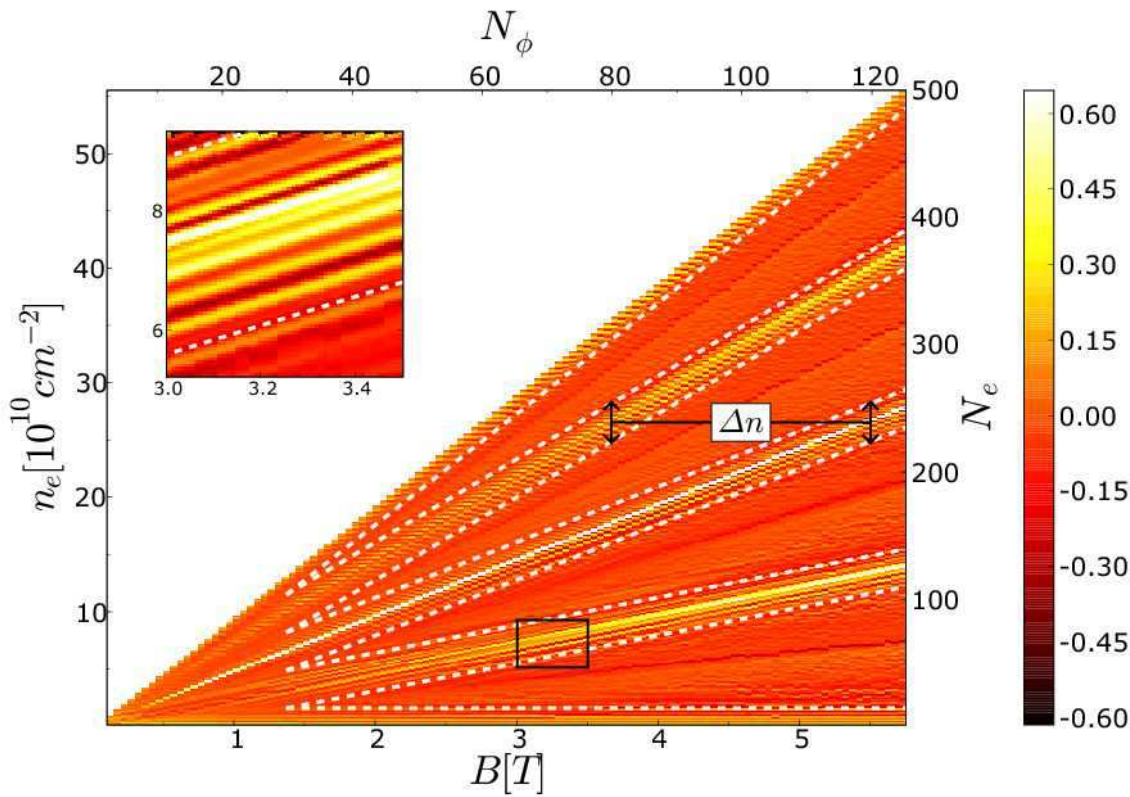}
  \caption{Inverse electronic compressibility $\kappa^{-1}$ for a
    HF-interacting system of size $L=300$nm with disorder strength
    $W/d^2 = 1.25$meV in the $(B,n_{\rm e})$-plane. The dotted lines show
    estimates based on a perfect screening condition (see text for
    details). The color scale spans two standard deviations around the
    average of $\kappa^{-1}$. The inset shows more detailed results for
    the region marked by a black rectangle.}
  \label{fig-int-compress-1}
\end{figure}

\begin{figure}[h]
  \center
  \includegraphics[width=\textwidth]{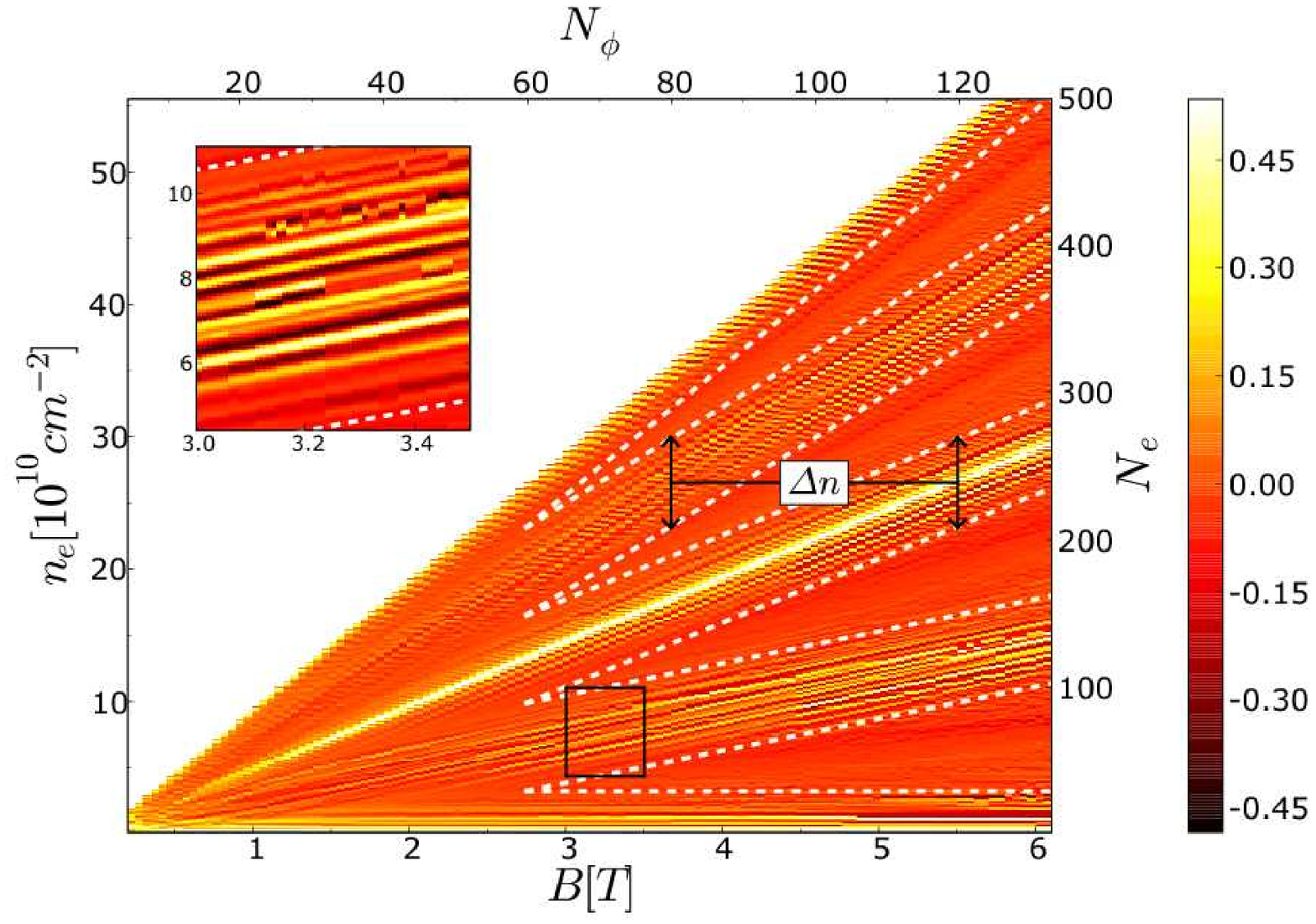}
  \caption{Inverse electronic compressibility as in Figure
    \ref{fig-int-compress-1} but with $W/d^2 = 2.5$meV.}
  \label{fig-int-compress-2}
\end{figure}

\begin{figure}[h]
  \center
  \includegraphics[width=\textwidth]{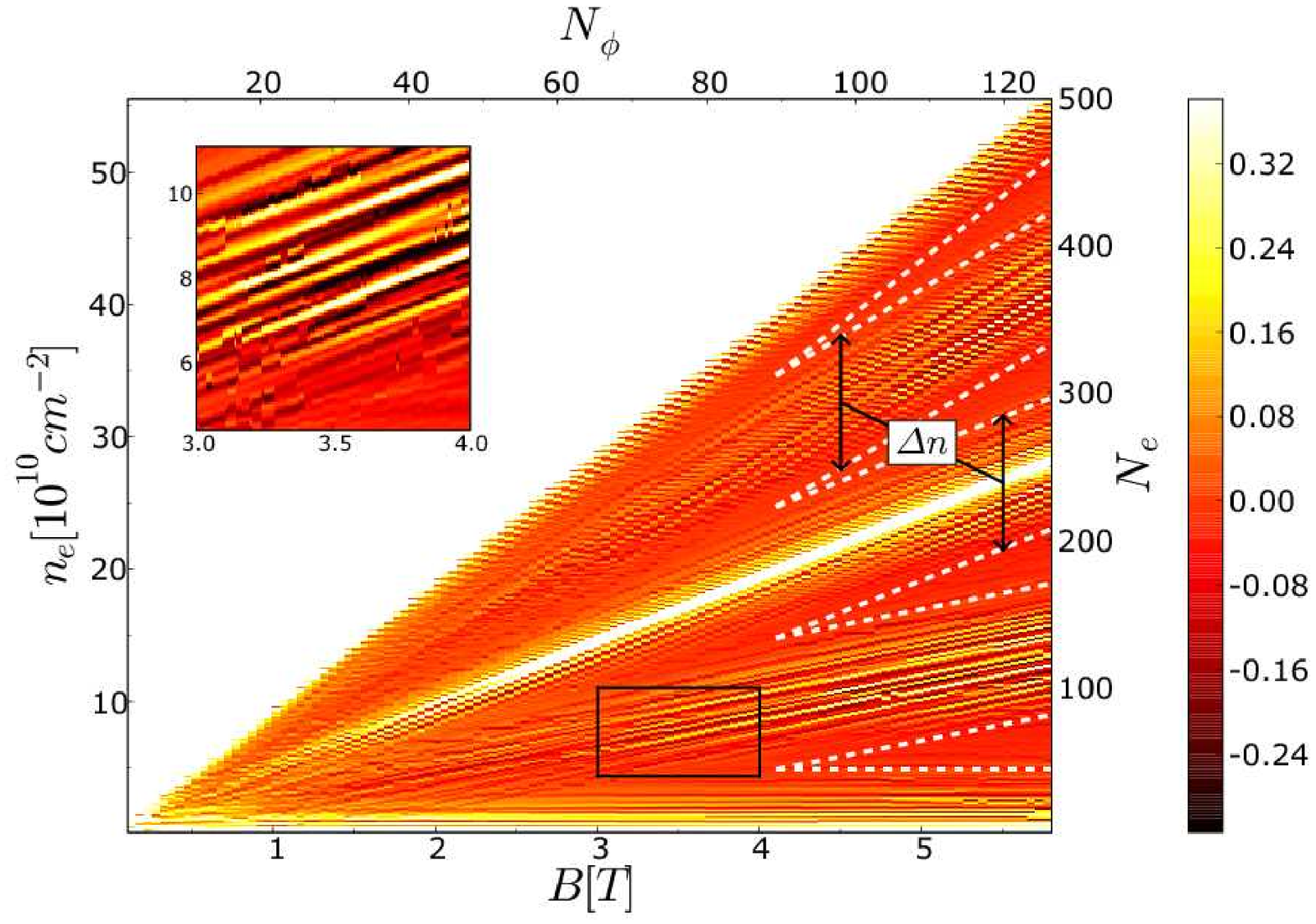}
  \caption{Inverse electronic compressibility as in Figure
    \ref{fig-int-compress-1} but with $W/d^2 = 3.75$meV.}
  \label{fig-int-compress-3}
\end{figure}

Figures \ref{fig-int-compress-1}, \ref{fig-int-compress-2}, and
\ref{fig-int-compress-3} show results in the $(B,n_{\rm e})$-plane for
an interacting system of size $L=300$nm with disorder strengths $W/d^2 =
1.25, 2.5$, and $3.75$meV, respectively, at fixed impurity range
$d=40$nm.
We observe that the exchange interaction results in an effective
$g$-factor substantially enhanced from its bare value
\cite{Jan69,NicHKW88,ManG95}, leading to a clear separation of the two
spin bands. This yields two additional strongly incompressible stripes
at $\nu=1$ and $\nu=3$, indicated by particularly high $\kappa^{-1}$
values. Quite different from the non-interacting case, we find that most
of the incompressible lines form groups which align parallel in the
$(B,n_{\rm e})$-plane along integer filling factors. Above a certain
minimal magnetic field, the width of these groups --- the number of the
lines --- is independent of the magnetic field and Landau level, forming
{\it incompressible} stripes of constant width around integer filling
factors. Overall, this behaviour is strikingly similar to the effects
observed in the experiments of \cite{IlaMTS04,MarIVS04}. Outside the
stripes, there is hardly any feature in the compressibilities except
directly at the QH plateau-to-plateau transitions at half-integer
fillings where a small increase in compressibility is discernible. In
these areas between incompressible stripes, the inverse compressibility
tends to have a very low or even negative value, which relates to a very
high or negative TDOS, respectively.  This effect has been observed
experimentally \cite{KraPS89,EisPW92} and is a signature of the exchange
interaction. From the proportionality between compressibility and the
screening length, we can conclude to observe strong overscreening in the
areas of negative compressibility. We attribute this to the tendency of
the HF-interacting 2DES to form a charge density wave
\cite{YosF79,CotM91}.
Furthermore, when comparing Figures \ref{fig-int-compress-1},
\ref{fig-int-compress-2}, and \ref{fig-int-compress-3} we find that the
width of the incompressible stripes increases with increasing disorder
strength $W/d^2$.

%The dotted white lines indicate regions where incompressible lines are
%absent and can be calculated from a perfect screening condition, given
%in the Appendix.

%%%%%%%%%%%%%%%%%%%%%%%%%%%%%%%%%%%%%%%%%%%%%%%%%%%%%%%%%%%%%%%%%%%%%%%%%%%
\section{Spatial distribution of electronic density and screening}
\label{sec-density}
%%%%%%%%%%%%%%%%%%%%%%%%%%%%%%%%%%%%%%%%%%%%%%%%%%%%%%%%%%%%%%%%%%%%%%%%%%%

\begin{figure}[h]
  \center
  \includegraphics[width=\textwidth]{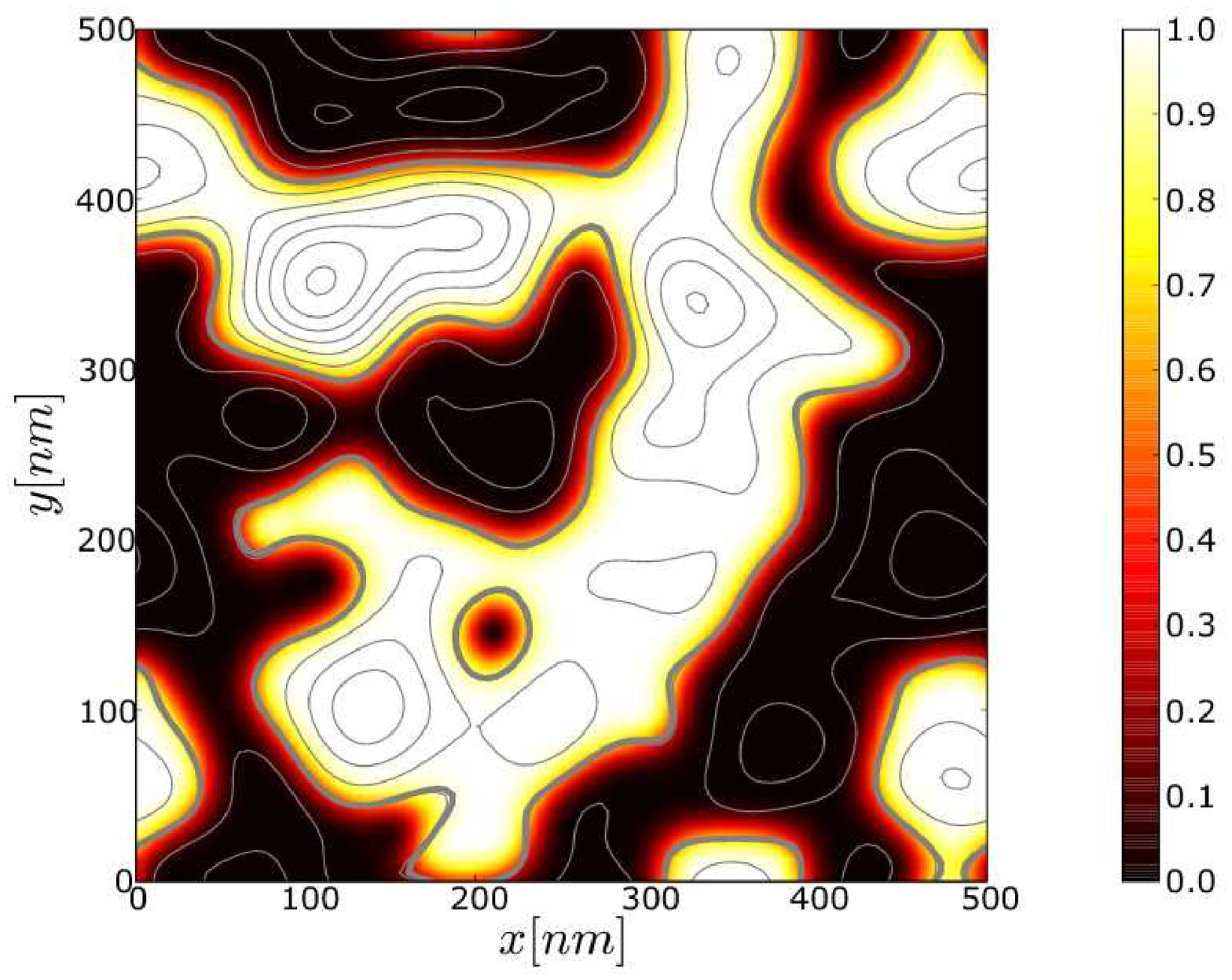}
  \caption{Spatial distribution of non-interacting electron density
    $n(\vec{r})/n_0$ at $B=4$T, $\gamma=0$ and $\nu=1/2$ as indicated by the
    color scale. Solid contour lines show the equipotential lines of the
    $\VD(\vec{r})$. The thick solid lines corresponds to $\epsilon_{\rm F}$. }
  \label{fig-density-non}
\end{figure}

\begin{figure}[h]
  \center
  \includegraphics[width=\textwidth]{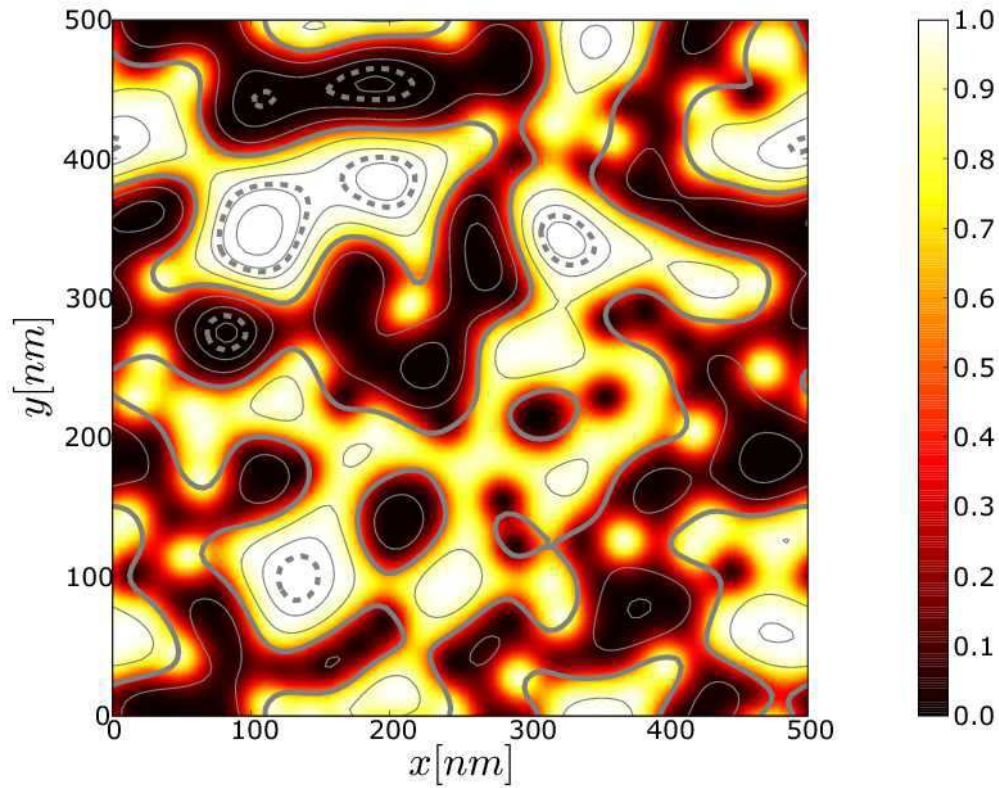}
  \caption{Spatial distribution of HF-interacting electron density
    $n(\vec{r})/n_0$ at $B=4$T, $\gamma=0.3$ and $\nu=1/2$ as indicated by the
    color scale. Contour lines show (\ref{eq-qV}). The broken lines
    indicate unscreenable (insulating) regions. The thick solid line shows
    $n_{\rm scr}(\vec{r}) = \bar{n}_{\rm scr} = 0$.}
  \label{fig-density-int-linear}
\end{figure}

The spatial distribution of the total electronic density
\begin{eqnarray}
n(\vec{r})
&=\sum_{\sigma}\sum_{\alpha=1}^{M} \left|\psi_{\alpha}^{\sigma}(\vec{r})\right|^2 \\
&= L^{-2}\sum_\sigma \sum_{n,k,n',k'}\sum_{\vec{q}} \mathbf{D}^{\sigma}_{n,k;n',k'} S_{n,k;n',k'}(\vec{q})
   \exp(-\i\q\r)
\end{eqnarray}
is readily calculated in our model. It details the screening mechanism
by providing direct insight into the interplay of disorder and interaction.
Let us start at the QH transition.
Figure \ref{fig-density-non} depicts the critical charge density at $\nu = 1/2$ for
a non-interacting system in units of $n_0$.
The contour lines show the impurity potential $V_{\rm I}(\vec{r})$ where
the critical energy $V_{\rm I}(\vec{r}) = \epsilon_{\rm F}$ is
highlighted by a thick line.  The charge density evidently behaves
according to the semiclassical approximation \cite{Huc95} and follows
the equipotential lines of $V_{\rm I}(\vec{r})$.
For the interacting case, however, we expect Thomas-Fermi screening
theory to apply \cite{Efr88a,Efr88b,Efr89a,Efr92}.  The electrostatic
potential of the charge density
\begin{equation}
\phi(\vec{r}) = \frac{e}{4\pi\epsilon\epsilon_0}\int d^2\vec{r}' \frac{n(\vec{r}') - \bar{n}}{|\vec{r}'-\vec{r}|}
\end{equation}
and the impurity potential $V_{\rm I}(\vec{r})$ form a screened
potential $ V_{\rm scr}(\vec{r}) = V_{\rm I}(\vec{r}) + e
\phi(\vec{r})$.  Here, $\bar{n}$ accounts for the positive background.
Since a flat screened potential is energetically most favourable, one
expects to find $V_{\rm scr}(\vec{r}) = \epsilon_{\rm F}$ for the case
of perfect screening. However, since fluctuations of the density,
$\delta n(\vec{r}) = n(\vec{r}) - \bar{n}$, are restricted between an
empty and a full Landau level, i.e.~$0 < \delta n(\vec{r}) < n_0$, the
screening is not always perfect but depends on the fluctuations in the
impurity potential as well as on the filling factor
\cite{Efr88a,Efr88b,Efr89a}.  The plane can be divided into fully
electron or hole depleted, insulating regions --- where $n(\vec{r}) = 0$
or $n(\vec{r}) = n_0$, respectively --- and metallic regions --- where
$n(\vec{r})$ lies in between. Depending on the filling factor, the
extent of those regions varies. Close to the band edge, insulating
regions dominate. Screening is highly non-linear and transport virtually
impossible. On the other hand, if disorder is weak enough, there exists
a finite range of filling factors in the centres of each band where
metallic regions cover most of the sample, percolate and render the
whole system metallic. The disorder is effectively screened and
transport greatly enhanced. In that case, the charge density $n_{\rm
  scr}(\vec{r})$ can be obtained by Fourier transforming the screened
potential. In 3D, this simply leads to the Laplace equation. For 2D,
however, one obtains \cite{WulGG88}
\begin{equation}
n_{\rm scr}(\vec{q}) = -\frac{2\epsilon\epsilon_0}{e^2}|\vec{q}|V_{\rm I}(\vec{q}) + \nu n_0 \delta_{q,0}, %\qquad (q \ne 0)
\label{eq-qV}
\end{equation}
where the $|\vec{q}| = 0$ term is "perfectly screened"
by the positive background and thus does not contribute to screening of
the impurity potential. In other words, in our model only the fluctuations
$\delta n(\vec{\r})$ are essential for screening.
Hence, in 2D, a perfectly screening charge density would obey
\begin{equation}
n_{\rm scr}(\vec{r})
= -\frac{4\pi\epsilon\epsilon_0}{e^2}\int d^2\vec{r}' \frac{\Delta_{\rm 2D} V_{\rm I}(\r')}{|\vec{r}-\vec{r}'|} + \nu n_0.
\label{eq-quasi-laplace}
\end{equation}
Clearly, the actual charge density is expected to deviate from $n_{\rm
  scr}(\vec{r})$ for several reasons. Firstly, the fluctuations of
$n(\vec{r})$ are restricted as discussed above. Secondly,
(\ref{eq-quasi-laplace}) is valid for the Hartree case only. Taking the
Fock contribution into account will introduce short wavelength
fluctuations due to the tendency for crystalization. However, we still
expect the charge density to follow (\ref{eq-quasi-laplace}) in the
limit of $|\vec{q}| \rightarrow 0$ \cite{screening}. Figure
\ref{fig-density-int-linear} shows results for the charge density of
interacting electrons at $\nu=1/2$. Broken lines indicate the regions
where $n_{\rm scr}(\vec{r})$ exceeds the range for $\delta n(\vec{r})$
either below or above, i.e.~areas that cannot be screened at all and
thus exhibit insulating behaviour.
\begin{figure}[h]
  \center
  \includegraphics[width=0.75\textwidth]{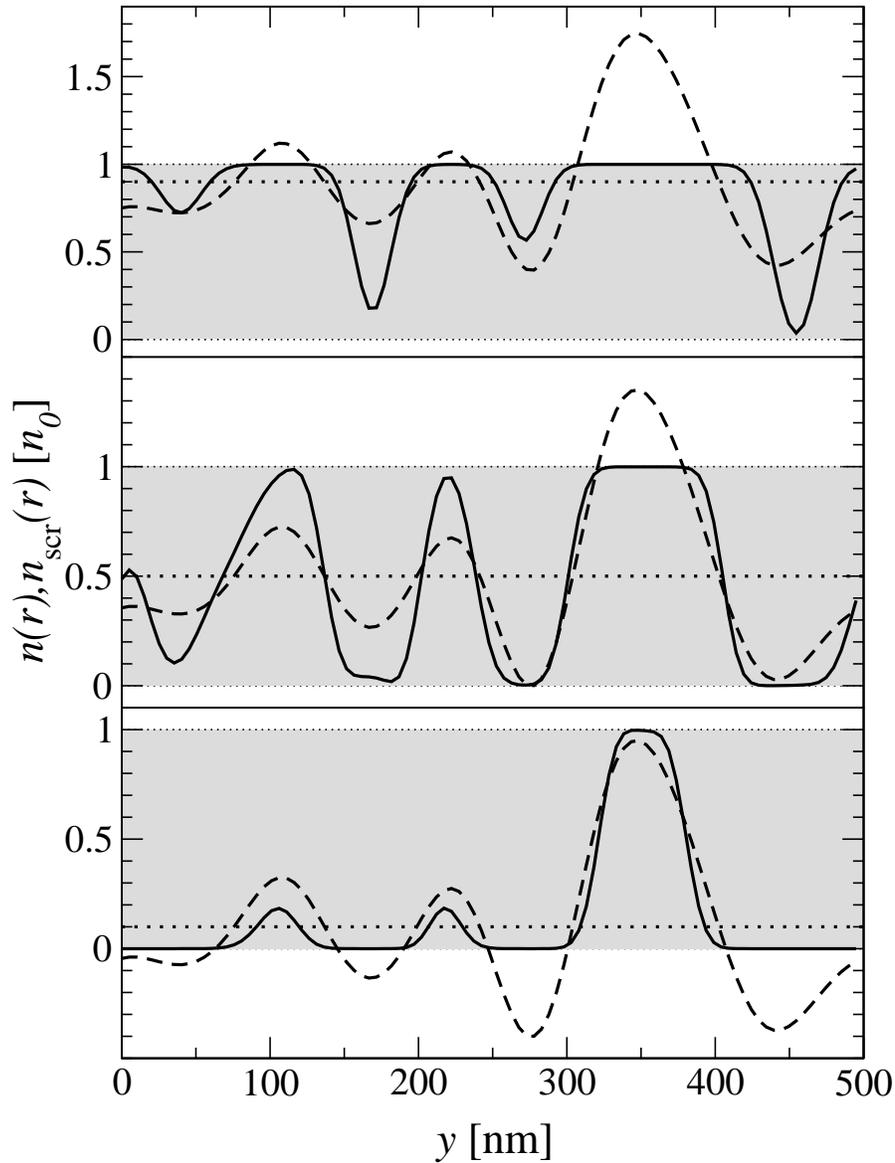}
  \caption{Cross-sections of the system of Figure
    \ref{fig-density-int-linear} at $x=100$nm and filling factors $\nu =
    0.1$, $\nu = 0.5$, and $\nu = 0.9$ (bottom to top). Full curves
    correspond to $n(\vec{r})$, broken lines show
    (\ref{eq-quasi-laplace}). A thick horizontal dotted line shows the
    average charge density, $\nu n_0$. The grey areas, bounded by thin
    dotted lines, indicate the complete band. }
  \label{fig-density-int-linear-cross-section}
\end{figure}
Otherwise, we find the charge density to follow $n_{\rm scr}(\vec{r})$
very closely. In this regime, the density is well described by
(\ref{eq-quasi-laplace}) and the screening is very effective. Metallic
regions dominate over insulating ones and transport is expected to be
good. In Figure \ref{fig-density-int-linear-cross-section} we depict
cross-sectional plots of $n(\vec{r})$ and $n_{\rm scr}(\vec{r})$ for the
sample of Figure \ref{fig-density-int-linear} at $x=100$nm and three
different filling factors, demonstrating the discussed effects again
very clearly.
\begin{figure}[h]
  \center
  \includegraphics[width=\textwidth]{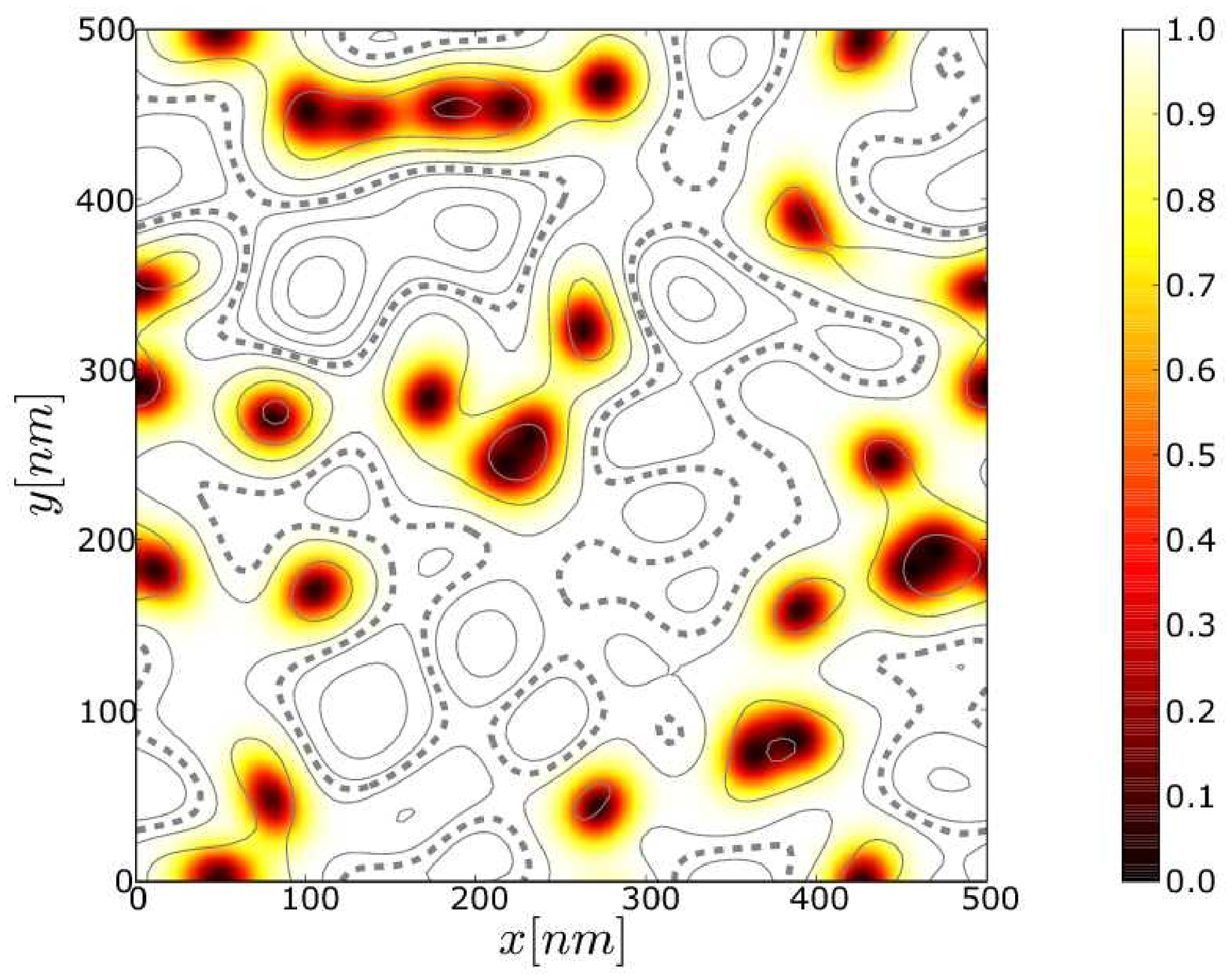}
  \caption{Spatial distribution of HF-interacting electron density
    $n(\vec{r})/n_0$ at $B=4$T, $\gamma=0.3$ and $\nu=0.9$ as indicated
    by a color scale. Solid contour lines show the equipotential lines
    of (\ref{eq-qV}). The broken lines indicate unscreenable
    (insulating) regions.}
  \label{fig-density-int-non-linear}
\end{figure}

%%%%%%%%%%%%%%%%%%%%%%%%%%%%%%%%%%%%%%%%%%%%%%%%%%%%%%%%%%%%%%%%%%%%%%%%%%%
\section{Breakdown of linear screening}
\label{sec-screening}

\begin{figure}[h]
  \center
  \includegraphics[width=\textwidth]{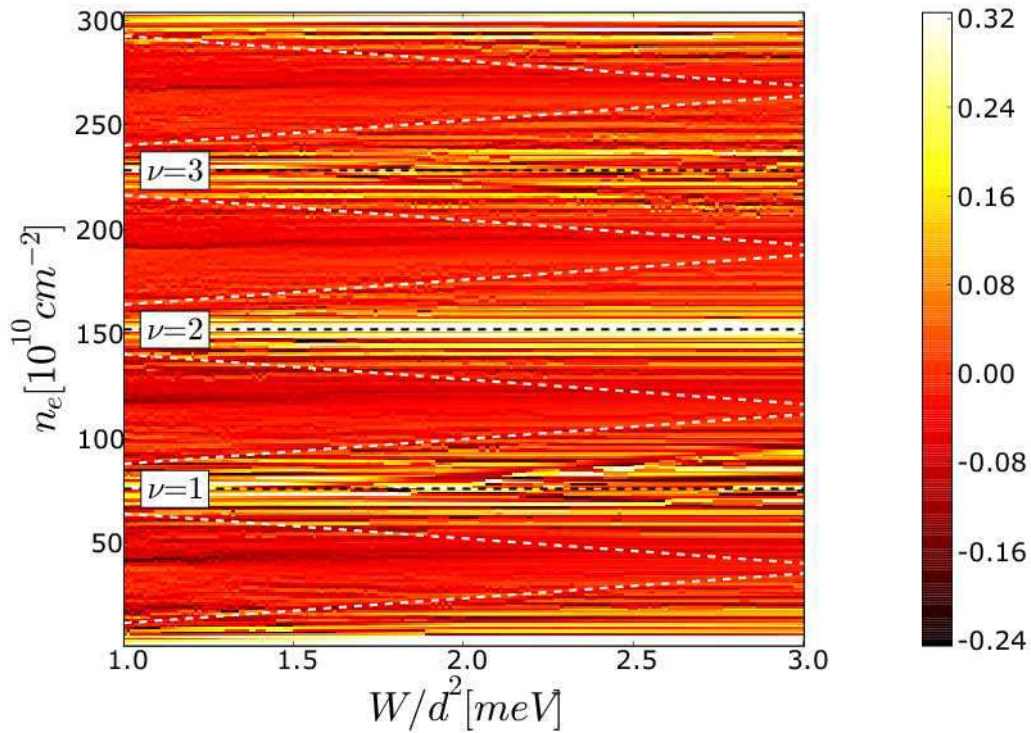}
  \caption{Inverse electronic compressibility $\kappa^{-1}$ for a
    HF-interacting system ($\gamma = 0.3$) of size $L=300$nm
    in the $(W/d^2,n_{\rm e})$-plane.
    Dashed white lines indicate expected boundaries of linear screening as calculated from (\ref{eq-deltan}),
    dashed black lines show integer filling factors.}
  \label{fig-V-n-plot}
\end{figure}

Thus far we have shown that our results can qualitatively reproduce the
structures observed in the $(B,n_{\rm e})$ plots of the compressibility.
We find stripes of constant width with very similar characteristics as
in the experiments. Furthermore, we show that within HF, the impurity
potential in the band centre can be quite effectively screened by the
charge density. Let us now turn our attention to the stripes.
The screening of the impurity potential is non-linear near the edges of
the Landau bands. Most of the sample is thus covered by insulating
regions where the Landau band is either completely depleted or filled.
Metallic behaviour is confined to small regions around potential
extrema, where electron or hole islands are formed. If additional
charge is introduced into the system, charging effects will govern the
spectrum. These charging effects will manifest themselves in jumps in
the compressibility as a function of charge density. The simple dot
model of \cite{PerC05} was able to account for this effect and even
demonstrated that charging events will take place along lines of integer
filling factor, in agreement with our calculations.
An estimation for the cross-over from linear to non-linear screening can
be found by very general considerations \cite{CooC93}. An insulating
island where $n(\vec{r}) = n_0$ is confined by the force of the impurity
potential, $\nabla V_{\rm I}(\vec{r})$, around its edge. Thereby, the
Coulomb interactions opposing this force making the edge of the full
region metallic. The size of the edges, i.e.~the size of the metallic
region, is then determined by the Coulomb force $n_0
e^2/2\pi\epsilon\epsilon_0$. Only if the Coulomb force acquires a
magnitude comparable to the typical confining potential force, the
metallic edges of the full islands will connect and dominate over the
insulating regions. The typical force of our impurity potential is given
by $ \langle|\nabla V_{\rm I}(\vec{r})|^2\rangle = {n_{\rm I} \langle
  {w_s}^2 \rangle_s}/{\pi d^4}$. We would like to remark that with
$N_{\rm I} = 288$, the expected standard deviation of ${\langle {w_s}^2
  \rangle_s}$ is $\sim 2\%$, which makes the typical force a reliable
characteristic of $V_{\rm I}$ for finite sample calculations. From
equating the typical force with the Coulomb force we can derive an
expression for the minimal required density $n_0$ which corresponds to
a minimum magnetic field $B_{\rm min} = n_0 h/e$ below which linear
screening breaks down for any density. Therefore, $n_0$ determines the
width of the charging stripes $\Delta n$ and we find
\begin{equation}
  \Delta n = n_0 =
  \frac{2\pi\epsilon \epsilon_0}{\gamma e^2}\sqrt{\langle|\nabla V_{\rm I}(\vec{r})|^2\rangle} =
  \frac{2\pi\epsilon \epsilon_0}{\gamma e^2}
  \sqrt{\frac{n_{\rm I}}{3\pi}}\frac{W}{d^2}.
  \label{eq-deltan}
\end{equation}
Note that $\Delta n$ is indeed independent of $B$ and $n_{\rm e}$ as
observed in the experiments. In Figures \ref{fig-int-compress-1},
\ref{fig-int-compress-2}, and \ref{fig-int-compress-3}, we have
indicated the breakdown of the linear screening regime by dashed white
lines. The points at which the lines merge indicate $B_{\rm min}$.
Evidently, (\ref{eq-deltan}) nicely estimates the widths of the observed
stripes for all three disorder configurations used. Furthermore, we have
tested the criterion for breakdown by plotting compressibilities as a
function of $n_{\rm e}$ and disorder strength $W/d^2$. Figure
\ref{fig-V-n-plot} shows the result for $W/d^2$ between $1$meV and
$3$meV at $B=3.5$T. The dashed white lines again indicate
(\ref{eq-deltan}). In order to confirm the dependence of the stripes on
the ratio between $W$ and $d^2$ only, the plot has been divided into two
regimes. Between $1$meV and $2$meV, we kept $d=40$nm as a constant and
varied $W$, and between $2$meV and $3$meV we kept $W/\mbox{nm}^2=3.2$meV
constant whilst varying $d$, accordingly. The results confirm
(\ref{eq-deltan}). Deviations from the expected behaviour can be
explained with the proximity to the disorder dominated regime for higher
values of $W/d^2$ where $B \simeq B_{\rm min}$. This regime is strongly
disorder dominated and charging effects become much less pronounced at
the band edges.

%%%%%%%%%%%%%%%%%%%%%%%%%%%%%%%%%%%%%%%%%%%%%%%%%%%%%%%%%%%%%%%%%%%%%%%%%%%
\section{Conclusions}
\label{sec-conclusions}
%%%%%%%%%%%%%%%%%%%%%%%%%%%%%%%%%%%%%%%%%%%%%%%%%%%%%%%%%%%%%%%%%%%%%%%%%%%

We have investigated numerically how electron-electron interactions
affect the localization properties of a 2DES under influence of a strong
perpendicular magnetic field. We therefore diagonalized the Hamiltonian
in a suitable basis and treated interactions as an effective mean field.

Our calculations reveal substantial differences in the electronic
compressibility between non-interacting and interacting systems when
viewed as a function of magnetic field and carrier density. For
interacting systems, we find strongly incompressible stripes of constant
width around integer filling factors. We show the dependence of the
width of the stripes on the disorder configuration and compute the width
based on a force balance argument. These results are in very good
agreement with recent imaging experiments. Moreover, we find strong
$g$-factor enhancement as well as negative compressibility in the band
centres, also consistent with experiments. We demonstrate that the
incompressible patterns can be attributed to non-linear screening
effects in the tails of the Landau bands. For magnetic fields
larger than $B_{\rm min}$, the effects of linear screening --- and hence
interactions --- dominate in the $(B,n_{\rm e})$-plane. Thus, our
results support the existence of a greater variety of transport regimes
due to electron-electron interactions in the integer quantum Hall
effect.

Similar compressibility patterns have also been observed around
fractional filling factors $\nu = 1/3, 2/5$, and $2/3$ \cite{MarIVS04}.
Energy gaps at fractional filling, e.g.~$\nu = p/(2p + 1)$, with $p$
being an integer, are a consequence of electron correlations which are
absent in HF approximation.  However, with the formal analogy
\cite{Jai03} between IQHE and FQHE put forward by the composite fermion
(CF) model \cite{Jai89,Jai00}, let us venture a few statements about
compressibility patterns around those fractional fillings. It is argued
that the FQHE can be regarded as a manifestation of the IQHE for CFs in
an effective magnetic field $B^* = B_{\nu} - B_{\nu=1/2}$
\cite{DuSTP93,StoTG99}.  If we pretend to have obtained results for CFs
in an $(B^*,n_{\rm CF})$ plane, a transformation back to electrons
yields an increase in the density of charging lines (per $n_{\rm e}$) by
a factor of $2p+1$.  Indeed, in the above mentioned experiment an
increase of $3$ has been found for $\nu = 1/3$. Furthermore, such a
transformation predicts a dependence of the width of the incompressible
stripes on the filling factor as well as a strong increase of $B_{\rm
  min}$ when fractional filling factors approach $\nu = 1/2$. This
remains yet to be explored.

%%%%%%%%%%%%%%%%%%%%%%%%%%%%%%%%%%%%%%%%%%%%%%%%%%%%%%%%%%%%%%%%%%%%%%%%%%%
\section*{Acknowledgements}
%%%%%%%%%%%%%%%%%%%%%%%%%%%%%%%%%%%%%%%%%%%%%%%%%%%%%%%%%%%%%%%%%%%%%%%%%%%
We gratefully acknowledge discussions with J.\ Chalker, N.R.~Cooper, A.\
Croy, B.\ Huckestein and A.\ Struck. Financial support has been provided
by EPSRC and the Deutsche Forschungsgemeinschaft (priority research area
``Quantum Hall Effect''). The computing facilities were provided by the
Centre for Scientific Computing of the University of Warwick with
support from Science Research Investment Fund grants as well as the
National Grid Service UK, where most of the numerical calculations have
been carried out.

\appendix

%%%%%%%%%%%%%%%%%%%%%%%%%%%%%%%%%%%%%%%%%%%%%%%%%%%%%%%%%%%%%%%%%%%%%%%%%%%
\section{Plane wave matrix elements and boundary conditions}
%%%%%%%%%%%%%%%%%%%%%%%%%%%%%%%%%%%%%%%%%%%%%%%%%%%%%%%%%%%%%%%%%%%%%%%%%%%
For $n \ge m$, the plane wave matrix elements $S_{n,k;n',k'}(\vec{q})$ read
\begin{eqnarray}
 \hspace{-2cm}
 \langle nk|\exp(\i\q\cdot\r)|mj\rangle &=
  \delta'_{q_y,k-j} \sqrt{\frac{2^n m!}{2^m n!}} \e{-\frac{\q^2}{4}+\frac{\i}{2}q_x(k+j)} \left(\frac{\i q_x - q_y}{2}\right)^{n-m} L_m^{n-m}\left(\frac{\q^2}{2}\right),
\end{eqnarray}
where $L_n^a(x)$ is the generalized Laguerre polynomial. For periodic
boundaries, the delta function is defined with modulus, i.e.\
\begin{eqnarray}
 \delta'_{a,b} &=
     \left\lbrace \begin{array}{ll}
      1 & \mbox{if} \quad \mbox{mod}(a-b,N_\phi) = 0, \\
      0 & \mbox{otherwise}. \\
     \end{array}\right.
\label{eq-DeltaFunction}
\end{eqnarray}
The periodic boundary conditions require careful treatment of the
overlap between Landau functions at opposite sites of the sample.  We
only take $|mj\rangle$ to be a replicated Landau function in one unit
cell to the right and one to the left of the sample, whereas $\langle
nk|$ remains in the base cell.  A check of the implementation can be
carried out by noting that a shift of the impurities by $L/N_\phi$ along
the $x$ or the $y$ direction should only shift the wave-functions,
$\psi_\alpha^\sigma(\vec{r})$, by the same value.
Finally we would like to remark that the complexity of summations
involving plane wave matrix elements can be greatly reduced by
neglecting terms where $\exp(-\vec{q}^2/2)<\varepsilon$. One can apply
the restriction $|q_x|^2 < \max(0,-2\ln(\epsilon)-q_y^2)$ to summations
over $q_x$, usually leading to a substantial reduction of complexity
even for $\varepsilon \sim 10^{-10}$.

%%%%%%%%%%%%%%%%%%%%%%%%%%%%%%%%%%%%%%%%%%%%%%%%%%%%%%%%%%%%%%%%%%%%%%%%%%%
\section{Bielectronic integrals}
%%%%%%%%%%%%%%%%%%%%%%%%%%%%%%%%%%%%%%%%%%%%%%%%%%%%%%%%%%%%%%%%%%%%%%%%%%%
Summations containing the bielectronic integrals
\begin{equation}
  G_{n,k;n',k'}^{m,l;m',l'} = \sum_{\vec{q}\ne0} v(\vec{q})
  S_{n,k;n',k'}(\vec{q}) %\langle nk|\e{\i\vec{r}\cdot\vec{q}}|n'k'\rangle
  S_{m,l;m',l'}(-\vec{q}) %\langle ml|\e{-\i\vec{r}\cdot\vec{q}}|m'l'\rangle,
\end{equation}
can be substantially simplified by virtue of the delta function
(\ref{eq-DeltaFunction}) contained in the plane wave matrix elements.
Ultimately, two of the summations in (\ref{eq-fockmatrix}) drop out. We
can replace $q_y$ by $k-k'$ and $l'$ by $l+k-k'$, and if $n \ge m$ as
well as $n' \ge m'$, we get
\begin{eqnarray}
  \hspace{-2cm}
  G_{n,k;n',k'}^{m,l;m',l'} =
  & \delta_{q_y,k-k'}\delta_{l',l+k-k'} \sum_{q_x} v(\q) \sqrt{\frac{2^{n_a}n_b!}{2^{n_b}n_a!}} \sqrt{\frac{2^{n_c}n_d!}{2^{n_d}n_c!}} \e{-\frac{\q^2}{2}+\i q_x (k'-l)} \times \\
  & \times \left(\frac{\i q_x-q_y}{2}\right)^{n_a-n_b}\left(\frac{-\i q_x+q_y}{2}\right)^{n_c-n_d}
  L_{n_b}^{n_a-n_b}\left(\frac{\q^2}{2}\right)L_{n_d}^{n_c-n_d}\left(\frac{\q^2}{2}\right).
\end{eqnarray}

%%%%%%%%%%%%%%%%%%%%%%%%%%%%%%%%%%%%%%%%%%%%%%%%%%%%%%%%%%%%%%%%%%%%%%%%%
% BibTeX users please use
%\bibliographystyle{prsty}
%\bibliography{bibliography/bibliograph}
%\bibliography{mybib}

\begin{thebibliography}{10}
\bibitem{overlapmatrix} In (\ref{eq-hfe}) we have omitted the overlap
  matrix for the molecular orbitals, which is simply the unity matrix
  for our Landau functions.
\bibitem{systemsize} When more realisticly Coulomb
  interaction between electrons and impurities is assumed,
  a logarithmic dependence on the sample size is found \cite{Efr88a}.
\bibitem{partratiocare} We emphasize that a proper use of participation
  numbers should not simply compare their values for different states at
  fixed system size as this can lead to wrong conclusions. Rather, their
  system size dependence must be studied \cite{EilRS98a}.
\bibitem{screening} Note that the matrix $\mathbf{\tilde{H}} \equiv
  \mathbf{V} + \mathbf{F_{\rm Hartree}}$, i.e.\ including only the
  disorder and Hartree terms, is simply $V_{\rm scr}(\vec{r})$ in the
  Landau basis. Thus an operation such as
  Tr$(\mathbf{\tilde{H}}\mathbf{\tilde{H}}^*)$ can be regarded as a
  measure for how well the disorder is screened \cite{Soh07}.
\end{thebibliography}

\end{document}